\documentclass[doublecol]{epl2} 
\usepackage{amsmath}
\usepackage{amssymb}
\title{Small Worlds in Space: Synchronization, Spatial and Relational Modularity}
\shorttitle{Small Worlds in Space: Synchronization, Spatial and Relational Modularity} 

\author{Markus Brede\inst{1} }
\shortauthor{M. Brede}

\institute{                    
  \inst{1} CSIRO Marine and Atmospheric Research, CSIRO Centre for Complex System Science, F C Pye Laboratory - GPO Box 3023, Clunies Ross Street
Canberra ACT 2601, Australia\\

}

\abstract{In this Letter we investigate networks that have been optimized to realize a trade-off between enhanced synchronization and cost of wire to connect the nodes in space. Analyzing the evolved arrangement of nodes in space and their corresponding network topology a class of small world networks characterized by spatial and network modularity is found. More precisely, for low cost of wire optimal configurations are characterized by a division of nodes into two spatial groups with maximum distance from each other, whereas network modularity is low. For high cost of wire, the nodes organize into several distinct groups in space that correspond to network modules connected on a ring. In between, spatially and relationally modular small world networks are found.  }

\begin{document}

\maketitle

\section{Introduction}

Synchronization phenomena occur in a diverse range of contexts in nature, engineering and society: cardiac pacemaker cells, neurons in the brain, fireflies that flash in unison, the power grid or consensus formation among people are just a few example applications from these fields. All of these are distributed systems embedded in space in which most couplings are local, but often also non-local long range couplings are present. Hence, most of these systems can be described as small world (SW) networks  \cite{SW}: nodes represent elementary units of the system such as neurons, fireflies, power stations or people and links in the network represent interactions that describe how the elementary units influence each other. Synchronization phenomena on SW and other complex networks have found much attention in the recent literature, see, e.g. \cite{synrev}, for a review. Moreover, in the system of coupled oscillators which we consider below, superior synchronization is essentially related to maximizing the second smallest eigenvalue of the graph Laplacian. This eigenvalue --the algebraic connectivity-- is an important invariant for undirected graphs and is, e.g., relevant for the analysis of the robustness of networks against node removal or for epidemic spreading \cite{Atay}.

In this research, one focus of interest has been to understand how the structure of the interaction network influences the dynamics of synchronization. Various network properties have been connected with superior synchronization and, even though at least full synchronization ultimately depends on fine details of the network structure \cite{Atay}, rough rules of thumb are that enhanced synchronization is positively correlated with short average pathlengths, disassortativity, large girths, and very homogeneous degree distributions. For instance, it has been found that synchrony optimal networks, termed `entangled' networks by the authors of \cite{Donetti}, are regular graphs.

Among the research on synchronization characteristics of networks, the problem of optimal synchronization with constraints has found much interest. For instance, various studies reported about the weighted link arrangement that gives optimal synchronization on a given network topology \cite{Motter2,Hwang1,Chavez,Nishi}, the optimal connection architecture of strongly connected directed networks \cite{MBS} or the structure of the `synchronization fitness landscape' in network space \cite{Nishi1}. Other work addressed the problem of optimal synchronization of non-identical units on sparse networks \cite{MB,Buzna}.

As another constraint, the interplay between synchronization properties of a network and its embedding in space has recently been discussed \cite{MBR}. In \cite{MBR} the spatial embedding of the network is represented as a constraint on the length of a wire needed to realize the connections of the coupling network. It is found that in situations where the cost of these spatial connections is high, the network that gives optimal synchronization is a SW, in which the length of the connections in space are distributed according to a power law $P(l)\propto l^{-\alpha}$, if $P(l)$ gives the probability that a randomly picked link connects oscillators of spatial distance $l$. The costlier the wire, the larger the exponent $\alpha$.

Modelling networked systems whose evolution is guided by considerations of function as well as spatial constraints has applications for a number of biological and technical systems. In the context of synchronization one may, for instance, think of neuronal networks for which synchronization plays an important role in the information processing. Further, neuronal networks are systems that have evolved and cost-considerations of long and short links have certainly played a role in shaping their structure. Interestingly, a number of recent studies that attempt to entangle the large-scale organization of functional brain networks \cite{Chaos} have revealed a modular small-world organization -- a network architecture that is very naturally explained by the model of spatially constrained synchrony-optimal networks discussed below.

In this Letter we extent the model of \cite{MBR} by another degree of freedom. Different to previous work we do not consider nodes as having a fixed position in space, but allow nodes to change locations. We do not only ask `How many links does a SW need to synchronize?', but also ask the question: `What is the relative arrangement of nodes in space that allows for superior synchronization with limited coupling?'. As we will discuss below, depending on constraints, this leads to yet another class of SWs which realize superior synchronization: relationally and spatially modular SWs.

\section{The Model}

In more detail, similar to \cite{Donetti,Motter2,Hwang1,Chavez,Nishi,MBS,Pecora} we study identical synchronization in systems composed of $N$ linearly coupled identical oscillators 
\begin{equation}
\label{E0}
 \dot{s_i}=f(s_i)+\sigma \sum_j A_{ij} [g(s_j)-g(s_i)].
\end{equation}
In the above equation $\dot{s_i}=f(s_i)$ gives the dynamics of the individual oscillators, while $A_{ij}$ is the adjacency matrix of the coupling network, $\sigma$ the coupling strength and the function $g$ describes the ``inner'' coupling of the oscillators. For simplicity we restrict the study to undirected networks, i.e. symmetrical coupling matrices $A_{ij}$. Depending on the details of the oscillators $f$, the coupling strength and architecture $\sigma A_{ij}$ and the coupling function $g$, in the system (\ref{E0}) identical synchronization can occur. Analysing the stability of the fully synchronized state $f(s)=0$ Pecora and Carroll \cite{Pecora} have derived a `master stability function` which relates the stability to the eigenvalues of the Laplacian matrix $G_{ij}=\delta_{ij} \sum_i A_{ii}-A_{ij}$ of the coupling network. Since we are interested in undirected graphs, the spectrum of $G$ is real. Assuming that the networks are connected there is exactly one zero mode (which corresponds to perturbations along the synchronization manifold) and the eigenvalues $\lambda_i$, $i=1,...,N$ of $G$ may be labelled in ascending order $0=\lambda_1<\lambda_2\leq ...\leq \lambda_N$. In Ref. \cite{Pecora} it has been shown that for a large class of oscillators the stability of the synchronized state is related to a small eigenratio $e=\lambda_N/\lambda_2$. A similar synchronization measure can be derived for the discrete analogue of Eq. (\ref{E0}), synchronization on coupled map lattices, cf. \cite{Atay}.

The eigenratio is thus a measure for the stability of the fully synchronized state. Its independence of the details of the oscillator system allows it to define a measure for the synchronization properties of a network. For this reason it has been widely adopted in various studies \cite{Atay,Donetti,Motter2,Hwang1,Chavez,Nishi,MBS,Nishi1,MBR,Pecora}, there being also some support that it even gives a rough measure for synchronization properties of systems of non-identical oscillators \cite{Chavez,MBS}.

To proceed, we introduce the spatial aspect of the system. We consider oscillators with locations $\Delta x_i$, $i=1,...,N$ which are distributed on a 1-dimensional space with periodic boundary conditions. The parameter $\Delta$ defines the length of a measurement unit on that space, which we set to $\Delta=1$ in the following considerations. Defining $D_\text{max}=\max_i x_i$ a distance metric may then be defined by $d(i,j)=\min (|x_i-x_j|, D_\text{max}-|x_i-x_j|)$, such that the amount of wire needed to connect the oscillators is given by $W=\sum_{i<j} A_{ij} d(i,j)$.

In the following, we are interested in network configurations and spatial node arrangements that allow for superior synchronization properties while minimizing the amount of wire needed to realize them, i.e. networks that minimize an `energy'
\begin{equation}
\label{E2}
 E=\beta W/N + (1-\beta) e.
\end{equation}
In Eq. (\ref{E2}) the parameter $\beta$ determines the relative desirability of both factors, superior synchronization properties (measured by the eigenratio $e$) and minimal cost of connections (measured by the amount of wire per node $W/N$). The formalism of investigating trade-offs in network formation in Eq. (\ref{E2}) is very similar to approaches in previous studies, e.g., \cite{MBR,Sole,Gopal}. As it stands, the solution to the problem (\ref{E2}) is trivial if the locations of the oscillators are not fixed in space: all oscillators would move to one location, thus allowing for full connectivity at no cost of wire. In a real-world situation constraints such as limited mobility of the nodes, physical barriers to movement, minimum space requirements of nodes or other functional reasons that limit nodes to certain parts of the space prevent this configuration. We model the sum of these `real-world' limitations as a constraint that the average spatial distance $D(x)=\sum_{i<j} d(i,j)$ between nodes is held constant during the minimization of (\ref{E2}). The formulation of the problem in this framework allows to study what relative arrangement of oscillators in space gives rise to optimal synchronization for least cost of wire.

\begin{figure*}[tbp]
 \begin{center}
 \includegraphics[width=.9\textwidth]{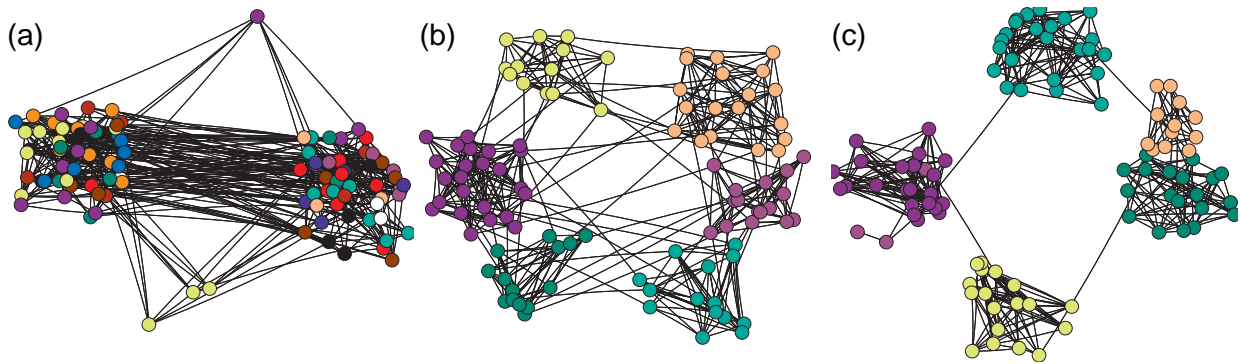}
 \caption{Examples of evolved networks for different trade-offs between cost of wire and desireability for superior synchronization: (a) $\beta=0.01$ (very low cost of wire), (b) $\beta=0.5$ (balanced costs for wire and synchronization), and (c) $\beta=0.01$ (very high cost of wire). The networks are of size $N=100$ and contain $L=400$ links. In the figure vertices have been colored according to the modules they belong to (modularities are $Q=.26$ for (a) and $Q=0.71$ and $Q=0.78$ for (b) and (c)). The spatial locations roughly correspond to the evolved spatial locations of the nodes during the optimization, however a random number was added to make vertices distinguishable.}
\label{fig.1}
\end{center}
 \end{figure*}

We continue by numerically constructing oscillator configurations that minimize $E(\beta)$ subject to $D(x)=\text{const.}$ for different trade-off parameters $\beta$. The numerical optimization scheme, which implements a variant of simulated annealing, is similar to schemes employed in previous studies, like, e.g., \cite{Donetti,Gopal,MBR,Sole}. It essentially consists of the following steps:

(i) Start with a configuration of oscillators that are evenly distributed on a 1-dimensional ring, i.e. $x^0_i=i$, $i=1,...,N$, which also defines the average spatial distance of the oscillators $D_0=D(x^0)$.  The coupling matrix in the initial condition is assumed to be given by an Erd\"os-R\'enyi random graph \cite{ER} with exactly $L$ links. 
(ii) Suggest either a rewiring or a location change for one or several oscillators. For rewiring suggestions, moves of randomly selected links to randomly selected `link vacancies' are suggested (provided that the new links do not introduce self-loops and that the new configuration is connected). In the case of a location change, a move of a randomly selected oscillator location $x_i \to x_i+\delta x_i$ (provided that $x_i+\delta x_i \geq 0$) and rescaling $x_i\to x_i D(x)/D(x+\delta x)$ of the oscillator locations such that $D(x)=D_0$ is suggested.
 (iii) Calculate the energy $E^\prime(\beta)$ of the modified configuration and accept if $E^\prime<E$ or with probability $p=\exp(-\nu (E^\prime-E))$ otherwise. As usual in simulated annealing procedures the `inverse-temperature parameter' $\nu$ is gradually increased during the optimization. If, according to the above rule, a suggested configuration is not accepted, the previous configuration is restored.
 (vi) Iterate steps 2 and 3 until no improvement in $E(\beta)$ could have been obtained for the last $10 L$ iteration steps. The motivation for the terminating condition is to stop the algorithm after every link or node location has been unsuccessfully tried to be modified several times.

It should be emphasized that the above optimization problem is a difficult non-linear problem and the numerical procedure does not ensure that a global optimum has been reached. However, we repeated the stochastic experiment for different initial conditions and the features about optimal configurations that we report below have been found to be robust.
\section{Results}

Before discussing how varying trade-offs between synchronizability and cost of wire influence the structure of the optimal network configurations, it is worthwhile to investigate the limiting cases $\beta=0$ and $\beta=1$ of our model. The case of $\beta=0$, no cost of wire, i.e. when the spatial arrangement of nodes becomes irrelevant, corresponds to the model of \cite{Donetti}, which we briefly discussed in the introduction. Importantly, optimal networks for $\beta=0$ are not modular and --since space is irrelevant-- the spatial arrangement of nodes in the optimal configuration is uniform. The optimal configuration for the other limiting case $\beta=1$, minimization of the cost of wire without regard for synchronization, is an arrangement of nodes into two distinct spatial clusters of nodes at maximum distance. Nodes within each of the two clusters are strongly interconnected, such that they also correspond to two network modules. These two modules are connected by exactly one link. Due to its community organization this is a network configuration with very poor synchronizability, cf. \cite{Arenas}.

\begin{figure}[tbp]
 \begin{center}
 \includegraphics[width=.45\textwidth]{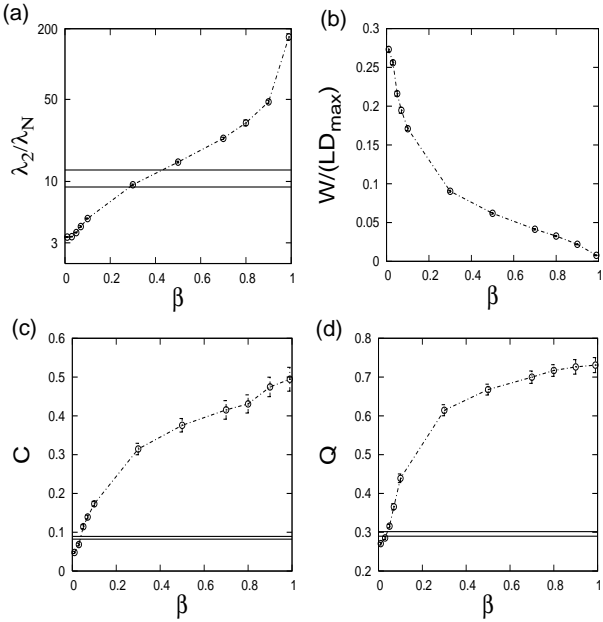}
 \caption{Dependence of some properties of the evolved networks and spatial arrangements on the trade-off parameter $\beta$: (a) eigenratio $e=\lambda_N/\lambda_2$, (b) average link length (measured in units of the maximum coordinate), (c) clustering coefficient $C$, (d) modularity $Q$. For reference, the horizontal lines indicate the range the respective quantities would assume for an Erd\"os-R\'enyi random graph whose nodes are uniformly distributed in space. In (b) the lines are omitted for scaling reasons, one has $W/(LD_\text{max}) \approx 0.5$.}
\label{fig.2}
 \end{center}
 \end{figure}

\begin{figure}[tbp]
 \begin{center}
 \includegraphics[width=.45\textwidth]{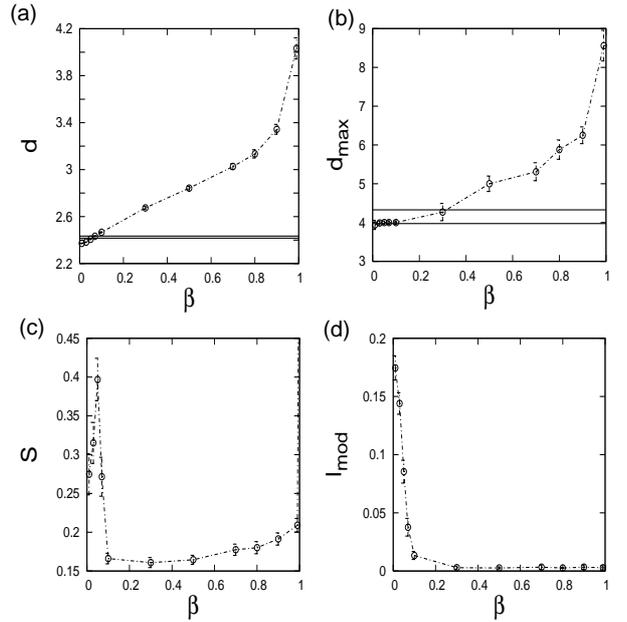}
 \caption{Dependence of (a) the average shortest pathlength $d$, (b) average diameter $d_\text{max}$, (c) spatial modularity $S$, and (d) average relative spatial distance between nodes in the same module in units of $D_\text{max}$ on the trade-off parameter $\beta$. For reference, the horizontal lines indicate the range the respective quantities would assume for an Erd\"os-R\'enyi random graph whose nodes are uniformly distributed in space. In (c) and (d) the lines are omitted for scaling reasons, one has $S<10^{-3}$ and $l_\text{mod}\approx .25$, respectively.}
\label{fig.2a}
 \end{center}
 \end{figure}

To proceed, Figure \ref{fig.1} shows some representative example networks of $N=100$ nodes and $L=400$ links evolved for three different values of the trade-off parameter $\beta$: an example when wire is inexpensive for $\beta=0.01$ (cf. Fig. \ref{fig.1}a), an example for when the cost of wire and desirability of superior synchronization properties are balanced for $\beta=0.5$ (cf. Fig. \ref{fig.1}b) and an example for a situation when wire is very expensive for $\beta=0.99$ (cf. Fig. \ref{fig.1}c). In the figures, nodes with spatially close locations are also drawn close to each other, even though distances are sometimes magnified or shrunk to make individual nodes distinguishable. The illustrations of the synchrony-optimized example networks strongly suggest the emergence of different types of modular network organizations that depend on the trade-off parameter $\beta$.

To explore the relational or network modularity, we have calculated the modularity $Q=\sum_m [L_m/L-(d_m/2L)^2]$ introduced in \cite{Girvan} for the networks. In the expression for $Q$ the index $m$ extends over all modules, $L_m$ denotes the number of links between nodes of module $m$, and $d_m$ is the sum of the degrees of all nodes in module $m$. Since the example networks investigated are relatively small we have analyzed the modules via extremal stochastic optimization \cite{Duch}, but have also tried other methods like \cite{Girvan}, which have robustly confirmed the visual expectation one gathers from the network plots in Fig. \ref{fig.1}, in which nodes belonging to the same network module are characterised by identical colouring.

Clearly, when wire is inexpensive and superior synchronization the dominant consideration, nodes assemble into two spatial clusters that are separated by the maximum spatial distance. In this case, however, the network- or relational structure of the coupling has very low modularity and the modules found by the algorithm appear uncorrelated with the spatial arrangement to the eye. 

When the demands for wire minimization and superior synchronization are balanced networks like the one visualized in Fig. \ref{fig.1}b emerge. The example network clearly decays into several distinct relational modules and already has a high network modularity of $Q=0.71$. Likewise, however, spatial clusters of nodes emerge in close correspondence with the network modules. Even though the modules are clearly distinct, they are still relatively strongly interconnected, mostly by long distance links.

The tendency towards stronger spatial and relational modularity continues when $\beta$ is further increased. For $\beta=0.99$, when wire economy is the main consideration in the optimization, very cohesive spatial and network modules can be discerned, cf. Fig. \ref{fig.1}c. Unlike as for the balanced case, these modules are connected to each other in a ring like arrangement by short links between spatial nearest  neighbour modules. 

For a more systematic investigation we constructed $R=100$ optimized network configurations for systematically varied trade-off parameters $\beta$. All networks investigated below are of size $N=400$ and have $L=400$ links. By displaying some key statistics of network arrangement and node arrangement in space, figures \ref{fig.2} and \ref{fig.2a} give an overview over the parameter space. By plotting the eigenratio $\lambda_2/\lambda_N$ and the average link length (in units of the maximum spatial coordinate $D_\text{max}$), panels (a) and (b) of Fig. \ref{fig.2} visualize the trade-off between superior synchronization properties and cost of wire. As expected, for small $\beta$ much wire can be used to achieve superior synchronizabilities, whereas for large $\beta$ costly wire allows for only few long links, poor synchronization behaviour being the consequence. To further classify the optimal network topologies, we investigated their degree distributions, degree variance, clustering coefficients (as defined in \cite{SW}), modularity, and the average shortest pathlengths and diameters. In agreement with \cite{Donetti}, we find that for the full range of $\beta$ parameters the degree distributions of the evolved networks are very narrow and the degree variances are significantly lower than for random graphs with the same number of nodes and links. However, as reported in \cite{MBR}, degree variances do not become exactly zero when $\beta>0$ and increase when $\beta$ is increased, i.e. when spatial constraints become more important.

More interestingly, panels (c) and (d) of Fig. \ref{fig.2} illustrate that the optimized networks become increasingly cliquish and modular when the relative cost of wire is increased. Thus, even for balanced cost of wire and desirability of superior synchronization, when synchronizablities in the range of Erd\"os-R\'enyi random graphs are attained, the network organization is characterized by a distinctly modular arrangement, cf. also Fig. \ref{fig.1}b. While for low $\beta$ the modules are connected by many long range links, these long range connections are increasingly thinned out when $\beta$ increases. As a consequence, while the modular arrangement becomes more and more distinct, diameter and shortest pathlengths of the optimized networks grow, cf. panels (a) and (b) of Fig. \ref{fig.2a}.

On the other hand, when wire is cheap, triangles are suppressed and no relational community organization with more cohesiveness than found in a random graph is detected, cf. Fig. \ref{fig.2}c,d. In this case the networks also become increasingly smaller, approaching the entangled net configurations discussed in \cite{Donetti}.

\begin{figure}[tbp]
 \begin{center}
 \includegraphics[width=.4\textwidth]{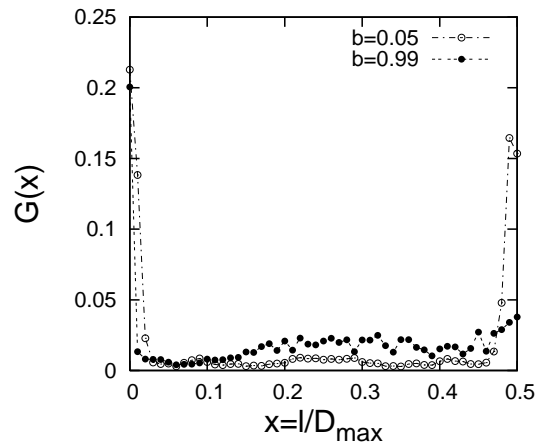}
 \caption{Spatial correlations: probabilities to have nodes at distance $[x,x+dx]$ from an average node for $\beta=0.05$ and $\beta=0.99$. Distances scaled in units of the maximum coordinate $D_\text{max}$.  }
\label{fig.3}
 \end{center}
 \end{figure}

As the network plots in Fig. \ref{fig.1} already show, the optimized configurations are not only characterized by a distinct network arrangement, but also have a characteristic arrangement of nodes in space. To classify the spatial arrangement of the nodes, we calculated a spatial correlation function $G(x)$ that measures the average probability of finding a node at spatial distance $x$ from an arbitrary node. Figure \ref{fig.3} displays plots of $G(x)$ for two typical spatial node arrangements found for $\beta=0.05$ and $\beta=0.99$. The presence of one or two distinct peaks in the function indicates a high degree of spatial clustering. While the two peaks comprising each roughly $N/2$ nodes for $\beta=0.05$ confirm the presence of two spatial clusters separated by the maximum spatial distance, the one peak that comprises about $N/6$ nodes for $\beta=0.99$ indicates the presence of several spatial clusters. In the multiple cluster configuration, individual clusters are not separated by a clearly defined typical spatial length scale, but tend to avoid each other (i.e. $G(x)$ has a broad minimum at ranges $x=0.02$ to $x\approx0.12$). The typical width of the peaks $\Delta x=0.01$ allows to define a more condensed spatial modularity measure $S=\int_0^{\Delta x} G(x)dx$, that quantifies the average number of nodes in the immediate spatial vicinity of an average node, thus defining the average relative size of a spatial module. Panel (c) of Fig. \ref{fig.2a} gives the dependence of $S$ on $\beta$ for the optimized network configurations. 
 
It is also of interest, whether spatial clusters overlap with network modules. To characterize this relationship, we calculated an overlap parameter $l_\text{mod}$ that measures the average spatial distance between nodes belonging to the same network module in units of the maximum coordinate $D_\text{max}$. If spatial and relational modules overlap completely, one expects to have $l_\text{mod}<\Delta x/D_\text{max}$, whereas $l_\text{mod}\approx 1/4$ if spatial and relational modules are completely uncorrelated. The dependence of $l_\text{mod}$ on $\beta$ is plotted in panel (d) of Fig. \ref{fig.2a}.

As discussed above, for $\beta=1$ a network configuration characterized by two spatial clusters corresponding to two network modules is optimal. In this case one has $S=1/2$ and $l_\text{mod}\approx 0$. Introducing a small consideration of synchronizability into the fitness function defined in Eq. (\ref{E2}) leads to a sharp drop in $S$, the network is immediately distributed into a number of smaller modules that comprise many less than $N/2$ nodes. In this process the overlap of spatial and relational module structures is preserved. This almost complete overlap persists over a wide range of $\beta$ parameters until around $\beta=0.07$. This value of $\beta$ marks a sharp transition in the structure of typical optimized configurations. Below $\beta=0.07$ two spatial clusters that do not overlap with network cliques are found.  In this situation the link length distributions are marked by two sharp peaks: very short links connecting nodes in the same spatial cluster and many links of maximum length connecting nodes pertaining to different spatial clusters.

For $\beta=0$, the cost of wire and spatial constraints become irrelevant and thus nodes in optimal network configurations are distributed uniformly in space. Accordingly, below the transition at $\beta=0.07$ the spatial modularity first sharply increases when the spatial two cluster configuration becomes established. However, after reaching a maximum at $\beta=0.05$ the spatial modularity declines again, approaching $S=\Delta x/D_\text{max}$ from above.

\section{Discussion and conclusions}
In this Letter we have numerically constructed spatially embedded optimal networks that realize a trade-off between superior synchronization properties and cost of wire. Different from all previous work on optimal synchronization, we focussed on the interplay between the relational- and spatial  organization of the optimized networks.

As the most interesting feature of our analysis we find that the optimal configurations are characterized by an interplay of  spatial clustering and network modularity. Essentially two parameter regimes, which are separated by a sharp transition, have been identified. When considerations of enhanced synchronizability outweigh requirements for the economy of wire ($0<\beta<0.07$), the optimal configurations are characterized by an arrangement of nodes into two spatial clusters with many long range connections between them, but also a remainder of short connections which link nodes within the same spatial cluster. However, the link arrangement is such that the spatial clusters do not correspond to relational (or network-) modules. When wire economy is important but not dominant for $0.07<\beta<1$, the optimal networks were found to decay into a number of clearly separated network modules. These network modules closely correspond to spatial clusters of nodes.

Our work has a number of interesting implications. First, it suggests a new explanation for the emergence of network modularity via the minimization of the cost of a wire when nodes are free to arrange themselves in space during the minimization procedure. A spatial arrangement of nodes in clusters serves to reduce the cost of wire: links between nodes in the same spatial cluster can essentially be introduced without or with very little cost. The interplay of this process with an additional mechanism which favours an unmodular network arrangement (in our case the demand for enhanced synchronization) can cause the breakup of large modules and result in a network arrangement characterized by the presence of a large number of small network modules. There is a variety of candidates for such processes, the simplest of which is probably the minimization of average shortest pathlengths, which has been considered in \cite{Gopal,Sole}.

Second, in the context of synchronization processes on networks, our work demonstrates that an additional constraint --minimization of the cost of a wire needed to connect nodes embedded in space-- which appears plausible in the context of biological or technical applications can cause network organizations to be `optimal', which have before been identified as suboptimal when synchronization properties are analyzed by purely investigating the structure of the coupling network. Dynamically, this modular structure is associated with the presence of different timescales for synchronization \cite{Arenas}. Our work suggests that oscillators which synchronize at the same timescale will also be located close to each other in space, a finding that may have applications in neurobiology \cite{Chaos}.

Third, it appears of interest that when considering nodes that are free to move in space, power laws in the link length distribution of optimal networks reported in \cite{MBR,Gopal} are replaced by bimodal link length distributions. Hence, an investigation of link length distributions of, e.g., biological networks may allow conclusions about the nature of evolutionary processes that shape the networks.

This research was undertaken on the NCI National Facility in Canberra, Australia, which is supported by the Australian Commonwealth Government. I thank A. Kalloniatis and F. Boschetti for helpful comments.

\end{document}